\documentclass[5p,times]{cas-dc}
\usepackage{amsmath,amsfonts}
\usepackage{amssymb}
\usepackage{pifont}
\usepackage{makecell}
\usepackage{algorithmic}
\usepackage{algorithm}
\usepackage{array}
\usepackage[caption=false,font=normalsize,labelfont=sf,textfont=sf]{subfig}
\usepackage{textcomp}
\usepackage{stfloats}
\usepackage{url}
\usepackage{verbatim}
\usepackage{graphicx}
\usepackage{cite}
\usepackage{array, caption, tabularx,  ragged2e,  booktabs}
\usepackage[numbers]{natbib}

\def\tsc#1{\csdef{#1}{\textsc{\lowercase{#1}}\xspace}}
\tsc{WGM}
\tsc{QE}
\tsc{EP}
\tsc{PMS}
\tsc{BEC}
\tsc{DE}

\begin{document}
\let\WriteBookmarks\relax
\def\floatpagepagefraction{1}
\def\textpagefraction{.001}

\title [mode = title]{A Lightweight and Secure PUF-Based Authentication and Key-exchange Protocol for IoT Devices}                      

%
\author[1]{Chandranshu Gupta}

\cormark[1]

\ead{chandranshu.gupta@iitjammu.ac.in}

\credit{Conceptualization of this study, Methodology, Software}

\affiliation[1]{organization={Indian Institute
of Technology Jammu}, 
    city={Jammu},
    postcode={181221}, 
    country={India}}

\author[1]{Gaurav Varshney}
\ead{ gaurav.varshney@iitjammu.ac.in}

\credit{Data curation, Writing - Original draft preparation}

\cortext[cor1]{Corresponding author}



\begin{abstract}
The Internet of Things (IoT) is rapidly becoming a common technology that will improve people's lives by seamlessly integrating into many facets of modern life and facilitating information sharing across platforms. Device Authentication and Key Exchange are major challenges for the IoT, which is essential to modern living. Many IoT sensor devices are placed in unprotected environments where they are vulnerable to physical attacks as well as common security risks due to their limited resources and weak self-protection capabilities. Large computational resource requirements for cryptographic primitives and heavy message transmission during Authentication make the existing methods like Public Key Cryptography and Identity-based Encryption (IBE) not suitable for these resource-constrained devices. Physical Unclonable Function (PUF) appears to offer a solid, practical, and economical security mechanism in place of typically sophisticated cryptosystems like PKI and IBE. PUF provides an unclonable and tamper-sensitive unique signature based on the PUF chip by using manufacturing process variability. Therefore, in this study, we propose a lightweight Authentication and Key Exchange protocol suitable for resource-constrained IoT devices that make use of lightweight cryptographic operations like bitwise XOR, hash function, and PUF to provide safe communication and a lightweight Authentication solution to thwart physical assaults.
Despite several studies employing the PUF to authenticate communication between IoT devices from the aforementioned security concerns to the authors' knowledge, existing solutions require intermediary verifier/gateway and/or internet capabilities by the IoT device to directly interact with a Server to authenticate itself and hence, are not scalable when the IoT device works on different technologies like Bluetooth Low Energy (BLE), Zigbee, etc. To address the aforementioned issue, we present a system in which the IoT device does not require a continuous active internet connection to communicate with the server in order to Authenticate itself. The results of a thorough security study are validated against adversarial attacks and PUF-modelling attacks. For formal security validation, the AVISPA verification tool is also used. Performance study recommends this protocol's lightweight characteristics. The proposed protocol's acceptability and defenses against various adversarial assaults are supported by a prototype developed with ESP32.
\end{abstract}



\maketitle
\section{Introduction}
{I}{nternet} of Things (IoT) has emerged as the Key technology for Smart Cities, Smart Homes, Remote Healthcare, Cyber-physical systems, etc. \cite{gubbi2013internet}. IoTs have created a network of ubiquitous sensing, communication, and actuation that is smoothly integrated into many aspects of contemporary daily life and allows for information exchange across platforms.

\begin{figure}[!t]
\centering
\includegraphics[width=0.35\textwidth,height=0.35\textwidth]{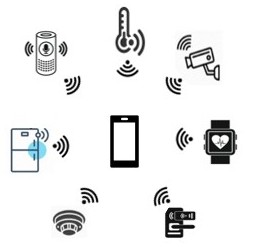}
\caption{General Application of IoT enabled Smart Home}
\label{home1}
\end{figure}

As shown in Figure \ref{home1}, IoT devices exchange a significant amount of data, including but not limited to various sensor readings such as temperature, blood pressure, heart rate, etc. Additionally, they transmit environmental data, such as the air quality index, pollution levels, etc. Location data, including GPS information, is also shared among IoT devices.
The devices utilized in an IoT environment usually produce significant amounts of security-sensitive data, making them highly alluring targets for attack as they are easily accessible and limited in resources \cite{meneghello2019iot}. 

IoTs are vulnerable to a wide range of serious risks, necessitating the establishment of security frameworks in a number of areas, including Authentication, Secrecy, Reliability, and Privacy. The typical attacks on IoTs include Side-Channel Attacks, Spoofing, Eavesdropping, Message Replay, Man-in-the-Middle (MitM), Cloning, Sleep Deprivation, Denial of Service (DoS), and Malicious code injection/forgery attacks \cite{sadhu2022internet}.
It's interesting to note that most of these assaults, if not all, are caused by security flaws in the existing Authentication protocols \cite{hassan2019current,nandy2019review}.

Authentication and safe Key Exchange are the initial lines of defense against phony endpoints that try to steal information \cite{nandy2019review}. Because Authentication is the foundation of all security measures, any weaknesses in the Authentication procedure would allow hackers to easily launch attacks, jeopardizing the system's security. This makes Authentication and Key management of IoT devices one of the most significant security concerns in the IoT ecosystem as they are resource-constrained and, therefore, unable to conduct computationally intensive cryptographic operations.

Public Key encryption has historically been used to manage Authentication and Key Exchange. The most widely used traditional method of Authentication and Key Exchange is Public Key Infrastructure (PKI) \cite{chatterjee2018building}. PKI is a system for creating, distributing, storing, and revoking digital certificates, as well as providing public-Key encryption \cite{garba2023lightcert4iots}. Certificate Authorities (CAs) are responsible for issuing PKI certificates that are digitally signed, linking the user's public Key to their identity. X.509 certificate format is typically used for certificate-based Authentication \cite{hallam2015x}. For IoT devices, the cost of certificate verification, certificate storage, and the complexity of verification present enormous hurdles. As a result, PKI is practically useless for IoT applications, where multiple devices are expected to connect. \cite{ting2017signcryption}. 

In order to address the challenges associated with PKI-based Authentication schemes for IoT devices, Identity-Based encryption (IBE) was introduced as a desirable alternative because it offers a method for creating public Keys from data that is already known to the public. The mechanism for generating, binding, and verifying public-private Key pairs without a digital signature is the primary distinction between PKI and IBE \cite{unal2021secure}. Typically, Identity-Based Encryption (IBE) systems obtain an entity's public Key from its identity string such as an email address or username. Also, a node's secret Keys are not connected to its physical identity. A private Key Generator (PKG) is necessary for traditional IBE in order to create private Keys for the nodes and conduct safe channel transfers. As a result, the Key Exchange is cumbersome and challenging for IoT application scalability and real-world implementation.
When handling many devices, the PKG would become a single point of failure and will lead to scalability constraints. Also, it introduces potential risks related to Key escrow, where the PKG can potentially decrypt any encrypted communication \cite{boneh2001identity}.

Contrary to traditional computer systems, the IoT is highly resource-constrained in terms of throughput, memory, processing power, and energy usage due to the widespread presence of resource-constrained devices, like sensors, actuators, and other Smart objects \cite{roy2022plake,jurcut2020security} as shown in Figure \ref{home1}. Therefore, traditional security measures (PKI and IBE) are insufficient to protect against the current IoT dangers. Hence, there is a need for lightweight Authentication mechanisms in terms of storage overhead, processing power, and energy consumption that can offer defense against various threats to the IoT ecosystem \cite{delvaux2015survey}.

In place of the traditionally sophisticated cryptosystems, Physical Unclonable Function (PUF) looks to be a robust, practical, and affordable security mechanism \cite{suh2007physical}. By taking advantage of manufacturing process diversity, PUF creates an unclonable and tamper-sensitive unique signature dependent on the PUF chip \cite{mahalat2021puf}. PUF relies on the idea of challenge-response pairs, in which each input challenge has its own corresponding response. Each PUF's set of challenge-response pairings is distinct. With enough basic PUF cells, each produced chip may generate countless different challenge-response pairs (CRPs). More crucially, since the PUF can generate the chip-specific secret via a unique response that is only generated upon request by providing a challenge, there is no need to keep a secret Key locally on a device. These characteristics make PUF an excellent choice for IoT devices with minimal resources \cite{sen2020puf}. 

Recently, lightweight PUF-based Authentication and Key Exchange protocols have received some attention in several IoT-related fields, including wireless sensor networks (WSNs) \cite{mahalat2021puf,mahalat2018puf}, Smart Homes \cite{cho2022secure}, Remote Healthcare \cite{masud2021lightweight}, Automotive Technology \cite{bansal2020lightweight}, and Internet of Drones \cite{alladi2020secauthuav}. Yet, the absence of formal verification, the need to maintain the CRP database at the Server end, and the use of complex cryptographic engines along with lightweight PUF instances render those protocols hostile and less suited for widespread implementation. 

\begin{figure}[!t]
\centering
\includegraphics[width=0.5\textwidth]{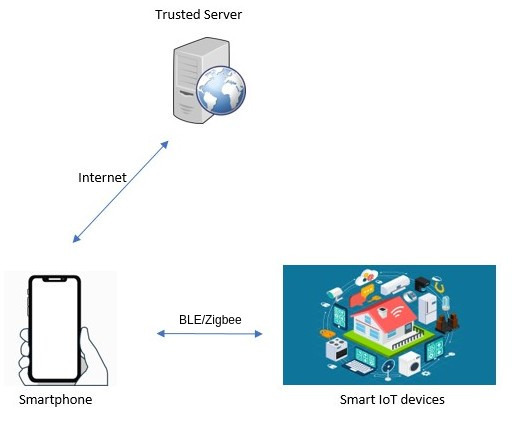}
\caption{Authentication scenario in our proposed solution}
\label{IoT1}
\end{figure}

\subsection{Motivation}
A growing subset of IoT, smart homes comprise a number of smart appliances that have microprocessor-based controllers built in. As shown in Figure \ref{home1}, the smart home is a system that makes use of a wireless sensor network (WSN) consisting of several smart devices like smart bulbs, washing machines, speakers, TVs, door locks, etc., that communicate with one another using IoT and can be controlled remotely via a Smartphone. Users may use a variety of home services in smart home environments, such as home monitoring, health monitoring, and assistance with everyday tasks. Despite the efforts made by researchers to secure smart home environments, securing smart home settings requires taking a number of security challenges into account.
In contexts where smart homes are used, entities communicate through open channels where an attacker can alter the messages being exchanged. This enables the attacker to carry out a number of security attacks, including message replay, device spoofing, user impersonation, and man-in-the-middle (MITM) attacks. 

In prior schemes, it is often assumed that IoT devices possess internet connectivity or rely on a gateway device as an intermediary for communication with a trusted server for Authentication and Key Exchange purposes. However, this assumption doesn't hold when devices utilize technologies such as BLE or Zigbee, where direct internet access may not be a prerequisite. 
Many existing solutions are even susceptible to Key escrow attacks as they depend on the Server for session Key generation, which is later used to encrypt IoT device communication.

Moreover, none of the proposed lightweight schemes tackle the scenario in which resource-constrained IoT devices, devoid of internet connectivity or a gateway and employing technologies like BLE, ZigBee, etc., within contexts such as Smart Homes or e-healthcare, need to undergo Authentication alongside a Client device—in our case, a Smartphone with internet capabilities—via the Server. In this specific Authentication phase, only the Client has the ability to engage directly with the Server, while the IoT device lacks that capability, as shown in Figure \ref{IoT1}.

These observations have served as the impetus for us to introduce a novel secure and lightweight Authentication and Key-Exchange scheme, leveraging Physical Unclonable Functions (PUFs), bitwise XOR, and secure hash function, ensuring the presence of essential security features within IoT-enabled smart home environments.

\subsection{Contribution}

We make the following major contribution to this paper:
\begin{itemize} 
    \item The IoT device does not need a real-time active Internet connection to communicate with the Server during the Authentication phase of our proposed protocol. Therefore, the proposed protocol is acceptable for devices that use BLE, Zigbee, RFID, and other wireless technologies. 
    \item Our proposed solution does not require the need to keep the CRPs database at the Server end. Moreover, no additional intermediary verifier/gateway is needed in our proposed protocol.
    \item We introduce a novel Authentication and Key-Exchange protocol, characterized by its lightweight and robust security, achieved by combining the PUF property with a secure Hash function and XOR operation.
    \item We conduct a formal security analysis of our proposed protocol using the Avispa validation tool. Furthermore, we scrutinize and compare security features, message overhead, and storage overhead with existing research in the domain. Additionally, we subject the proposed protocol to informal security analysis.
    \item We offer a Proof-of-Concept and assess the effectiveness of our proposed protocol through implementation on a real-time experimental testbed using ESP32.
\end{itemize} 

The remaining sections of this paper are organized as follows. Section II delves into related works, Section III provides preliminaries related to the paper's background, Section IV thoroughly describes our Lightweight and Secure PUF-Based Authentication and Key-Exchange Protocol designed for IoT devices, Section V conducts security analysis both informally and formally, Section VI presents testbed experiments with cryptographic primitives and comprehensive overhead comparisons with related schemes, and, finally, Section VIII concludes the paper.

\section{Related Works}
Numerous studies on IoT authentication and key exchange have been done in recent years. We will discuss them in this section.

\subsection{Public Key-based schemes}
PKI has traditionally been used for mutual Authentication and Key Exchange. In the IoT ecosystem, where we have billions of IoT devices, PKI suffers major challenges, including the overhead required for storing and exchanging these certificates and managing the certificate revocation process. When high levels of security are needed, expensive PKI is an appealing choice, but it is not the recommended primitive for IoT device Authentication \cite{ting2017signcryption}.

For IoT, Forsby et al. \cite{forsby2018lightweight} designed an X.509 profiled certificate by removing irrelevant data and compressing the rest using CBOR encoding. The size of conventional X.509 certificates was reduced by about 37\% as a result of similar work by Kwon et al. \cite{kwon2019lightcert}. PKIoT certificates, introduced in \cite{marino2019pkiot}, work on the principle that the resource-constrained devices only possess the link to the original certificate.

For BLE device Authentication, Gupta and Varshney \cite{gupta2023improved} developed a BLE profiled certificate for BLE devices. Compared to conventional X.509 Certificates, the proposed. The suggested certificate profile minimizes energy usage and certificate size when compared to traditional X.509 Certificates.

Even with these profiled certificates, the issue of distributing, verifying, and revoking these certificates is a major concern that prevents the use of PKI for IoT devices. 

To overcome the shortcomings of PKI, IBE-based schemes have been used. IBE uses ID numbers, email addresses, and phone numbers as users' unique public Keys. This attribute of a cryptographic system simplifies certificate generation and deployment without a certification authority. IBE systems were independently proposed by Boneh and Franklin \cite{boneh2001identity} and Cocks \cite{cocks2001identity}, both based on bilinear maps of particular elliptic curves. Bentajer et al. \cite{bentajer2018cs} presented CS-IBE  based on the original BF-IBE method \cite{boyen2007general}. Its foundation is the first Type-1 pairing, IBE, which has been revealed to have security flaws \cite{galbraith2008pairings}. It also relies on a single, centralized Key management authority, which is a single point of failure and could compromise privacy if a CSP is curious. Tan et al. \cite{tan2019enhancement} provide a non-pairing-based IBE system that does not enable multiple PKGs. In circumstances with a single PKG, the PKG can passively decode all messages while computing a Client's Private Key. To counter the problems caused by using a single PKG, Unal et al. \cite{unal2021secure} proposed SCSS-SAKKE-IBE, which employs Type-3 pairings and a fast pairing-based IBE scheme following the scheme in \cite{okano2020implementation}. Additionally, by utilizing the \cite{kate2010distributed} method, SCSS supports multiple distributed PKGs.

Although IBE-based schemes are beneficial when compared to PKI-based schemes since they don't have to deal with the overhead of handling certificates but IBE-based schemes use Bilinear pairings that require computationally demanding procedures, which is not practical for IoT devices with finite resources like memory, processing power, and energy. Bilinear pairings  cause significant overhead and performance reduction in such systems.

\subsection{PUF based schemes}
Recently, many studies have been conducted with the goal of creating PUF-based Authentication protocols for IoT applications. 
Many schemes that implement strong PUF for device Authentication have been put forth in the past.
Hashing the input challenge and the response is an idea by Gassend et al. \cite{gassend2008controlled}. This arrangement, however, necessitates hardware-expensive hashing and error correction logic in the device, rendering it highly infeasible for low-cost platforms. Additionally, the device needs to receive raw helper data from the Server in order to stabilize the noisy PUF responses. As a result, the PUF is vulnerable to attacks that concentrate on side-channel data \cite{becker2015pitfalls}.
Yu et al. \cite{yu2016lockdown} propose limiting the  CRPs., where only the Server or trustworthy entity can authorize new CRPs. This method enables just 10,000 Authentication cycles, making it unsuitable for applications requiring long device lifetimes. Hussain et al. \cite{hussain2018shaip} suggested a secure hamming distance-based mutual Authentication system using weak intrinsic PUFs and limitless Authentication cycles. This protocol's 487ms integrated processor delay makes it unsuitable for real-time Authentication. The protocol doesn't establish Keys either.

The PUF-based protocols in \cite{mahalat2018puf,masud2021lightweight,alladi2020secauthuav,huang2017puf,aman2017mutual,aman2020privacy} use the symmetric Key cryptosystem, which lowers the PKI's Key management overhead. In the context of wireless sensor networks (WSN), PUF-based procedures for Key Exchange and Authentication have been proposed, as seen in \cite{mahalat2021puf,mahalat2018puf,huang2017puf}. But implementing \cite{mahalat2021puf} necessitates substantial hardware upgrades for the IoT nodes. For a large-scale heterogeneous IoT network, the employment of multiple PUF-Challenge Response Pairs (CRPs) for Authentication leads to substantial database overhead on the Server side, as highlighted in the findings presented in \cite{mahalat2018puf}. Furthermore, as described in \cite{huang2017puf}, sensor nodes necessitate the storage of group Key factors within their own physical memory to facilitate the Authentication process. A secure protocol for intelligent Healthcare systems was described in \cite{masud2021lightweight}. The protocol, however, is slow when communicating between different nodes. Roy et al. in \cite{roy2021puf} put forth a lightweight PUF-based Authentication scheme for vehicle technology intercommunication. The protocol, however, is dependent on another protocol to fend off replay attempts that increase computation overhead.

In \cite{chatterjee2018building}, integration of PUF technology with certificate-less identity-based encryption is proposed. In this scheme, a verifier node acts as a proxy, facilitating the Authentication of two IoT nodes. Additionally, the Authentication process requires the involvement of a Security Association Provider (SAP) responsible for managing the Authentication aid information. In addition, the verifier must use an NVM to save a secret hash Key, as required by the protocol. Because of this constraint, the basic idea behind PUF—that it would prevent the secret from being stored in local memory—has been rendered useless.

In \cite{roy2022plake}, authors make use of lightweight XOR and hash operations along with random numbers and PUF CRP to authenticate the devices via the help of the Server.
Despite the fact that the implemented protocol is lightweight, it is vulnerable to key escrow attack and requires the IoT device to communicate with the Server using an active Internet connection directly and hence won't be suitable for devices that use BLE, Zigbee, RFID, and other wireless technologies.

In order to provide mutual Authentication across IoT nodes without resorting to local CRP storage, the research published in \cite{li2020provably} relies on combining PUF technology with PKC based on ECC. In this approach, following registration in a secure environment with the Server, each IoT node is provided with its distinct public and private Key pair. Running the protocol necessitates multiplying points on an elliptic curve sixteen times, significantly raising the procedure's computing cost.

As discussed above, in the existing PUF-based Authentication schemes,
\begin{itemize}
    \item  In most PUF-based solutions \cite{mahalat2018puf,masud2021lightweight,alladi2020secauthuav,roy2022plake,chatterjee2018building} employing lightweight cryptography, the IoT device requires an active internet and/or gateway during their Authentication phase for their scheme to work and thus are not suitable for devices using BLE, Zigbee, etc.
    \item \cite{li2020provably,chatterjee2018building} are not lightweight in terms of cryptography due to their usage of elliptic curve cryptography. 
    \item \cite{alladi2020secauthuav,roy2022plake} are prone to key escrow/insider attack.
\end{itemize}

Hence, in the context of this research paper, we have endeavored to address the issues mentioned previously.
\begin{itemize}
    \item  In our proposed scheme, the IoT device doesn't require active internet capabilities and/or a gateway device to communicate with the server to prove its authenticity. Only the Client device has internet capabilities.
    \item The lightweight cryptographic operations Bitwise XOR and Hash are used in our proposed method. 
    \item  Our proposed method is not vulnerable to key escrow/insider attack. Furthermore, just one CRP is required to be saved within the Server per IoT device, making mathematical modeling of PUF difficult.
\end{itemize}

\section{PRELIMINARY}
This section provides an overview of the preliminaries to enhance the paper's readability.

\subsection{Physical Unclonable Function (PUF)}
Gassend et al. \cite{gassend2008controlled} introduced the concept of PUF in 2002.
A PUF is a mathematical model that describes the behavior of a physical system using a collection of input and output pairs, denoted by  $(\mathbf{X}, \mathbf{Y})$. A PUF can be represented by the probabilistic function $f: \mathbf{X} \rightarrow \mathbf{Y}$, where $\mathbf{X}$ is the set of input challenges and $\mathbf{Y}$ is the set of output responses \cite{shamsoshoara2020survey}.

In a PUF, the input challenges $\mathbf{X}$ are typically n-dimensional binary vectors, i.e.,  $\mathbf{X} = (x_1, x_2, \ldots, x_n)$. Similarly, $\mathbf{Y}$ are binary vectors of length $m$, i.e.,  $\mathbf{Y} = (y_1, y_2, \ldots, y_m)$.

The function $f$ represents the physical properties of the system. It is a complex mapping between input challenges and output responses, involving a number of mathematical operations and functions. For instance, the PUF function $f$ can be described using the equation below:

\begin{equation}
\mathbf{Y} = f(\mathbf{X}) = \Phi(\Theta(\mathbf{X})),
\end{equation}

where  $\Theta(\cdot)$ represents an internal set of transformations and operations conducted on the input challenges $\mathbf{X}$, and $\Phi(\cdot)$ represents the final output mapping or response generation function.

Depending on the PUF architecture, the internal transformations $\Theta(\cdot)$ may include bitwise operations, modular arithmetic, error correction codes, or other mathematical operations. To generate the final output responses $\mathbf{Y}$, the function $\Phi(\cdot)$ may entail additional operations such as thresholding, nonlinear mappings, or error correction mechanisms.

PUFs have an inherent security characteristic known as "unclonability." A PUF is deemed unclonable if it is computationally impossible to determine the internal properties or reproduce the same behavior using the input-output pairs $(\mathbf{X}, \mathbf{Y})$.

The incapacity to accurately model or replicate the physical variations and their effect on the internal transformations and response generation function underlies the unclonability of a PUF. It requires precise knowledge of the physical properties and manufacturing flaws of the system, which is difficult to acquire.

The security of the PUF against cloning attacks and unauthorized access is guaranteed by the uniqueness of challenges and responses and the incapacity to reproduce the system's behavior.

PUFs leverage the distinctive differences added during the device's manufacture to obtain a fingerprint exclusive to the device. The device's threshold voltage, read/write times, leakage current, etc., are only a few of the particular characteristics that are monitored when an external stimulus is delivered.
When a parameter of a device is measured for the first time, the measurement is referred to as an original response for a particular input stimulus or a particular location in memory, referred to as a challenge, used to achieve this measurement. Both of these measurements are kept in the Server. A reaction occurs when the exact same parameter is assessed once again when the exact same external stimulus is present. This combination of challenges and answers, known as the Challenge Response combination (CRP), is often compared to one another in order to confirm the device's identification \cite{shamsoshoara2020survey}.
PUFs can be roughly divided into "strong PUFs" and "weak PUFs" according to the total number of CRPs they can generate. Weak PUFs are typically utilized for secret Key generation due to less number of CRPs they can generate and because of their great stability and reproducibility. Whereas in strong PUFs, a lot of CRPs are present, and hence, they are typically used for Authentication purposes.

To evaluate the usage of PUF, two statistical measures of intra-distance and inter-distance that, respectively, demonstrate the reproducibility and uniqueness of PUF may be used:
\begin{itemize}
    \item ``Intra-distance: the distance between two unique responses to the same PUF challenge measured'' \cite{bautista2016survey}
    \item ``Inter-distance: the distance between two replies of two separate PUFs to a given challenge.'' \cite{bautista2016survey}
\end{itemize}

\begin{figure}[!t]
\centering
\includegraphics[width=0.5\textwidth]{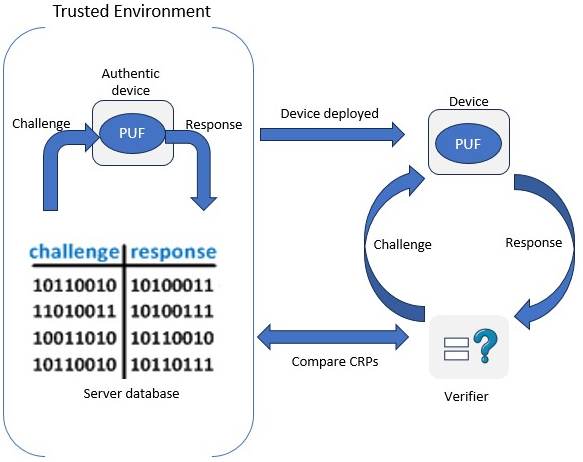}
\caption{Authentication scheme using PUF CRPs}
\label{CRP}
\end{figure}

The distinct CRPs generated by a device are what PUF-based security methods are dependent on \cite{korenda2018secret}. To utilize a PUF device with any cryptographic technique, it must first be registered with the Server during the enrollment phase, wherein a trusted secure environment set of challenges and their corresponding responses will be stored inside the Server database. The same set of challenges for the Client's PUF is used by the Server during the Authentication procedure to extract the corresponding responses, as shown in Figure \ref{CRP}. If the CRPs generated by the device match the stored CRPs inside the Server database then the device is authenticated.

\subsection{Protocol Assumptions}
Here, we go through the network model, which consists of three entities: an IoT device, a smartphone or laptop, and a server, as well as the protocol's threat assumptions.
\subsubsection{Network Model}
It is assumed that the resource-constrained IoT devices used in our system have an inbuilt PUF and the capability to perform both the XOR operation and cryptographic hash functions.

The Client device (in our case, a Smartphone) is not resource-constrained and is considered powerful enough to perform heavy cryptographic operations like ECDHE. It can perform hash operations and do a TLS handshake with the Server.

The Server possesses sufficient computational capabilities to manage a substantial array of IoT devices and Client devices while also performing resource-intensive cryptographic tasks such as ECDHE and secure hash. It also stores the authenticating PUF CRPs in a safe location. It is secure from the reach of enemies. It can perform a TLS handshake with the Client device
and let multiple Client devices log into the system using a username and password.

\subsubsection{Threat Assumptions}
In our proposed protocol, we assume that an attacker possesses access to the communication channel, allowing them not only to eavesdrop but also to actively attempt data corruption by introducing malicious code. Moreover, it is assumed that the embedded PUF's CRPs are an implicit characteristic of the underlying IoT device, and hence, an attacker can't gain access to it. Finally, we presume that only while utilizing the suggested protocol can the PUF of an IoT device that has registered with the server for authentication be accessible.
\subsection{Design Goals}
In this section, we delve into the design objectives of the proposed Authentication and Key Exchange Protocol based on PUF.

\begin{itemize}
    \item \textit{Maintaining only one PUF-CRP exclusively within the Server while ensuring resistance against model-building attacks:} For Authentication, the suggested protocol only needs a single PUF-CRP to be stored securely in the database of the Server. IoT devices only need to hold the PUF; they don't need to hold the CRPs. This is done to make sure that no attacker can use the CRPs to clone the PUF instances mathematically.
    \item \textit{No explicit secret Key storing:} IoT devices won't have explicit Key storage; instead, each IoT data node will have a PUF instance embedded to provide each device a distinct identity.
    \item \textit{Centralized control and scalability:} The proposed protocol relies only on the Server to add and control IoT devices and the Client device. Depending on its processing capability and available resources, the Server may add as many IoT devices and Client devices to its network.
    \item \textit{No dependence on Internet and Gateway device:} The IoT device does not need to be connected to the Internet in order to connect to the Server and complete the Authentication procedure. It can communicate with the Client device, which will speak to the Server and relay messages from IoT devices to the Server.
\end{itemize}

\section{Proposed Scheme}

\begin{table}[!t]
\caption{NOTATIONS\label{tab:table4}}

\centering
\begin{tabular}{|c||c|}
\hline
Symbol & Description\\
\hline
$Id_{i}$ & Identity of the device i\\
\hline
$C_{i}$ & PUF challenge for ith iteration \\
\hline
$R_{i}$ & PUF response to $C_{i}$ \\
\hline
$T_{i}$ (i=1,2,3,...,n) & Random Numbers \\
\hline
$N_{i}$ & Random Nonces of the device i\\
\hline
$f_{PUF}$ & PUF function\\
\hline
$\oplus$ & Cryptographic XOR operation\\
\hline
H(.) & Cryptographic hash function\\
\hline
\end{tabular}
\end{table}

A Lightweight and Secure PUF-Based Authentication and Key exchange protocol for Internet of Things devices is described in depth in this section. Table \ref{tab:table4} provides a brief explanation of the symbols used throughout this paper. There are two stages in our proposed protocol. 
\begin{itemize}
    \item Enrollment phase: During the enrollment phase, the IoT device and the Client device have to register with the trusted Server.
    \item Authentication and Key Exchange phase: During this phase, the Authentication and Key Exchange between the IoT device and the Client device is done.
\end{itemize}

\subsection{Enrollment and Registration phase}
\begin{figure}[!t]
\centering
\includegraphics[width=0.5\textwidth]{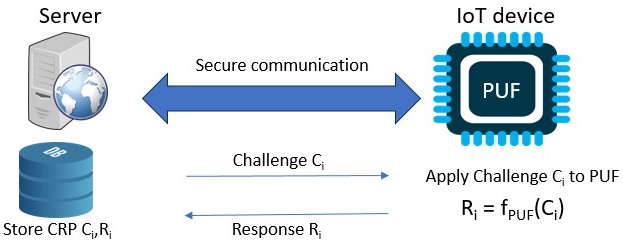}
\caption{Enrollment process of IoT device}
\label{enrol}
\end{figure}

The enrollment phase is performed in a secure environment, where an attacker can't intercept the data being transferred as shown in Figure \ref{enrol}. Here the trusted Server generates a random challenge ($C_{p}$) and sends this challenge to the IoT device having PUF embedded in it. The IoT device will then apply this challenge in its PUF and produce a response ($R_{p}$) corresponding to the challenge. The response is then sent back to the Server, and the Server stores the PUF CRP ($C_{p}$,$R_{p}$) inside its secure database for the IoT device ($Id_{p}$). This step is only performed once for an individual IoT device before the device is deployed/shipped.

After the enrollment phase is completed, the IoT device can be deployed, and the Authentication phase can be started. 

The Client device will also register itself to the Server. The registration step of the Client device is different from the IoT device as it is a resource-rich device having internet capabilities and doesn't have PUF embedded in it. The Client device will initially sign (over a TLS channel) into the Server using a Username ($U_{c}$) and Password ($P_{c}$). The Server stores the hash of the username, password, and the Client device identity ($Id_{c}$) in its database and generates an alias identity ($Id_{c'}$) for the Client device.

$Id_{c'}$ = H($U_{c}$,$P_{c}$,$Id_{c}$).

After the Client device has registered with the Server, the Client device is ready for the Authentication phase.

\subsection{Authentication and Key Exchange phase}
Upon successful enrollment of the IoT device and registration of the Client device within the Server system, they become prepared for communication. However, prior to initiating communication, a verification of their authenticity through the Server is essential. The forthcoming section outlines the specific steps involved in the Authentication phase.

\begin{figure*}[!t]
\centering
\includegraphics[width=\textwidth]{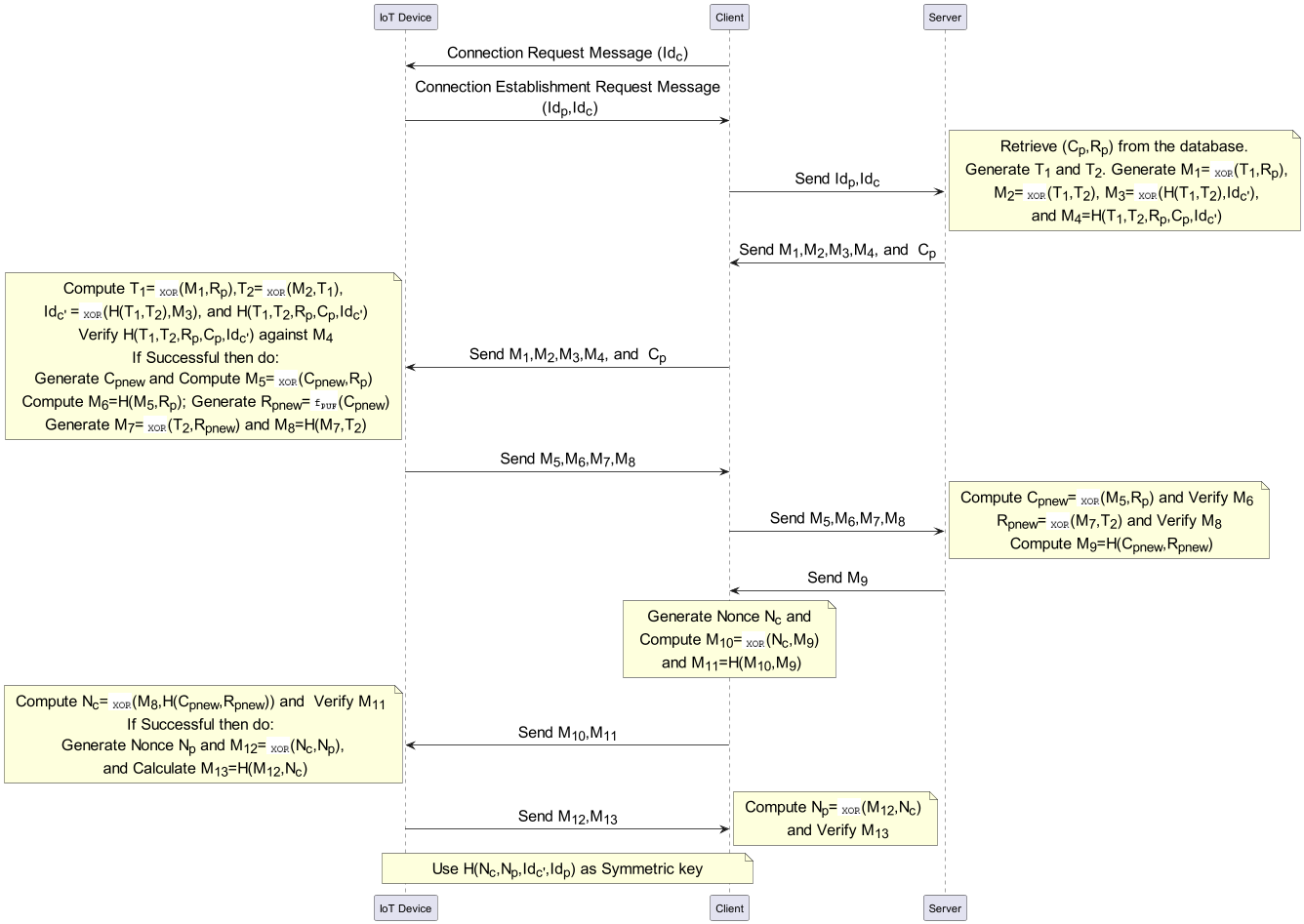}
\caption{Authentication and Key Exchange phase}
\label{IoT}
\end{figure*}

\begin{enumerate}
    \item The Client device sends the IoT device a connection request message containing $Id_{c}$. After receiving the request, the IoT device sends a connection establishment request message to the Client device containing its identification and the IoT device's identification that was received ($Id_{p}$, $Id_{c}$). The Client device then sends this connection establishment request message to the Server.
    \item Subsequently, the Server retrieves the CRP associated with the PUF of the IoT device from its database and generates two random numbers denoted as $T_{1}$ and $T_{2}$. The Server then produces $M_{1}$, $M_{2}$, and $M_{3}$ by, respectively, executing the XOR of $T_{1}$ with $R_{p}$, $T_{1}$ with $T_{2}$, and H($T_{1}$,$T_{2}$) with $Id_{c'}$. Finally, the Server generates $M_{4}$ by performing the H($T_{1}$,$T_{2}$,$R_{p}$,$C_{p}$,$Id_{c'}$). The Server transmits to the Client device the messages ($M_{1}$, $M_{2}$, $M_{3}$, $M_{4}$, $C_{p}$).
    \item The Client device then forwards the messages
    ($M_{1}$, $M_{2}$, $M_{3}$, $M_{4}$, $C_{p}$) received from the Server to the IoT device. Upon receiving the messages from the Client device, the IoT device will first input the $C_{p}$ to its inbuilt PUF and collect the matching response $R_{p}$. The IoT device then extracts $T_{1}$ from the received message $M_{1}$ via $R_{p}$ and then extracts $T_{2}$ from the message $M_{2}$ via $T_{1}$. The IoT device then performs H($T_{1}$, $T_{2}$) and decrypts $Id_{c'}$ from the message $M_{3}$ using H($T_{1}$, $T_{2}$). Finally, it verifies the $M_{4}$ message it received by performing H($T_{1}$,$T_{2}$,$R_{p}$,$C_{p}$,$Id_{c'}$) and matching it with $M_{4}$ for integrity checking.
    \item The IoT device will then create a new Challenge $C_{pnew}$ by computing a different permutation of the stable addresses of SRAM cells (discussed in detail in Section 6). Then, the IoT device will compute a message $M_{5}$ by performing the XOR of $C_{pnew}$ with $R_{p}$. After this, the IoT device will apply the new challenge $C_{pnew}$ to its PUF and generate a new response $R_{pnew}$. IoT device will then generate a new message $M_{6}$ by performing H($M_{5}$, $R_{p}$). Now using $T_{2}$, the IoT device then produce message $M_{7}$ by XORing $T_{2}$ with $R_{pnew}$. Following that, the IoT device will generate message $M_{8}$ by performing H($M_{7}$,$T_{2}$) and send to the Client messages $M_{5}$,$M_{6}$,$M_{7}$, and $M_{8}$ which the Client will then forward to the Server.
    \item The Server then fetches the new Challenge $C_{pnew}$ by XORing $M_{5}$ and $R_{p}$ and performs H($M_{5}$,$R_{p}$) matching it with $M_{6}$ for integrity checking. After that, the server fetches $R_{pnew}$ via XOR of $T_{2}$ and $M_{7}$ and verifies $M_{8}$. The old CRP ($C_{p}$ $R_{p}$) is removed from the database of the Server, and the new CRP ($C_{pnew}$ and $R_{pnew}$) is stored in the database. Finally the Server generates message $M_{9}$ by performing the operation H($C_{pnew}$,$R_{pnew}$) and sends $M_{9}$ to the Client device. 
    \item The Client generates a Nonce $N_{c}$ and produces message $M_{10}$ by computing XOR of $N_{c}$ with  $M_{9}$. It also produces message $M_{11}$ by performing \newline H($M_{10}$,$M_{9}$). Finally, the Client will send $M_{10}$ and $M_{11}$ to the IoT device. 
    \item The IoT device, after receiving $M_{10}$ and $M_{11}$ will fetch $N_{c}$ using H($C_{pnew}$,$R_{pnew}$) via XOR operation and also verify the the message $M_{11}$. It will then generate its own Nonce $N_{p}$ and produce a message $M_{12}$ by XORing $N_{p}$ with $N_{c}$. It will also generate message $M_{13}$ by computing H($M_{12}$,$N_{c}$) and send $M_{12}$,$M_{13}$ to the Client. The Client will fetch $N_{p}$ from $M_{12}$ using $N_{c}$ and verify the integrity of message $M_{13}$. At last, both the devices will use H($N_{c}$,$N_{p}$,$Id_{c'}$,$ID_{p}$) as the shared secret session Key for communication. This shared secret Key gets updated whenever a new session is established between the IoT device and a Client.
\end{enumerate}

\section{Security Analysis}
To verify the suggested protocol, we have carried out a thorough security analysis in this section.

\subsection{Protection against Mathematical Modeling}

To establish a mathematical model for the PUF, several researchers have explored machine learning methods such as Support Vector Machines (SVM) and Logistic Regression (LR). In order to train the system to comprehend the challenge-response behavior and accurately replicate the PUF, it is necessary to use a subset of Challenge-Response Pairs (CRPs) from a specific PUF instance. After training, the model may reasonably anticipate a response $R_{p}$ to any challenge $C_{p}$. 

Gathering the Challenge-Response Pairs (CRPs) from a specific PUF instance involves obtaining physical access to the IoT device and then generating responses for randomly generated challenges. To mitigate this risk, it has been stipulated that access to the PUF of a deployed IoT device is restricted solely to the authentication phase within the proposed protocol. The other way by which an attacker can access the CRPs is by eavesdropping on the communication channel. This is also prevented in our proposed protocol by using XOR operation, as shown below.
\begin{align*}
&M_{1} \oplus M_{2} = \left ({T_{1} \oplus R_{p}}\right) \oplus \left ({T_{1} \oplus T_{2}}\right) = R_{p} \oplus T_{2}\tag{1}\\
&M_{1} \oplus M_{3} = \left ({T_{1} \oplus R_{p}}\right) \oplus \left ({H(T_{1},T_{2}) \oplus Id_{c'}}\right)\tag{2}\\
&M_{2} \oplus M_{3} = \left ({T_{1} \oplus T_{2}}\right) \oplus \left ({H(T_{1},T_{2}) \oplus Id_{c'}}\right)\tag{3}\\
&eqn \left ({1}\right) \oplus eqn \left ({2}\right) = \left ({T_{1} \oplus T_{2}}\right) \oplus \left ({H(T_{1},T_{2}) \oplus Id_{c'}}\right)\tag{4}\\
&eqn \left ({1}\right) \oplus eqn \left ({3}\right) = \left ({T_{1} \oplus R_{p}}\right) \oplus \left ({H(T_{1},T_{2}) \oplus Id_{c'}}\right)\tag{5}\\
&eqn \left ({2}\right) \oplus eqn \left ({3}\right) = \left ({T_{2} \oplus R_{p}}\right) \oplus \left ({H(T_{1},T_{2}) \oplus Id_{c'}}\right)\tag{6}\\
&eqn \left ({4}\right) \oplus eqn \left ({5}\right) = eqn({1})\tag{7}\\
&eqn \left ({4}\right) \oplus eqn \left ({6}\right) = M_{1}\tag{8}\\
&eqn \left ({5}\right) \oplus eqn \left ({6}\right) = M_{2}\tag{9}
\end{align*} 

Looking at the cryptanalysis findings shown in the equations above, we can infer that accessing the challenge $C_{p}$ does not expose its matching response $R_{p}$, demonstrating the proposed protocol's resilience to adversary attempts to obtain the secret Key.

There is another way in which the CRPs of a particular PUF instance can be collected. Consider a scenario where a malicious Client device tries to connect to the Server to gather the CRP of an IoT device. Due to the singular storage of a CRP within the Server's database, the collection process permits only a single CRP to be acquired at a given moment. Moreover, this acquisition occurs in the form of XOR messages, as elucidated in the earlier cryptanalysis. Hence a malicious Client can't collect the CRPs of an IoT device and is therefore secure against mathematical modeling.

\begin{figure}[!t]
\centering
\includegraphics[width=0.3\textwidth]{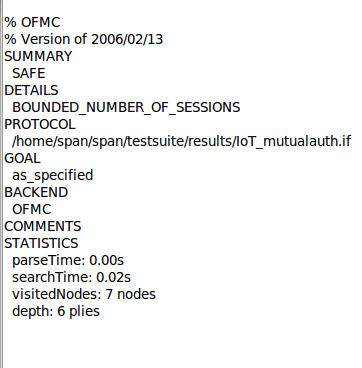}
\caption{AVISPA simulation results using OFMC (Single session)}
\label{ofmc2}
\end{figure}

\begin{figure}[!t]
\centering
\includegraphics[width=0.3\textwidth]{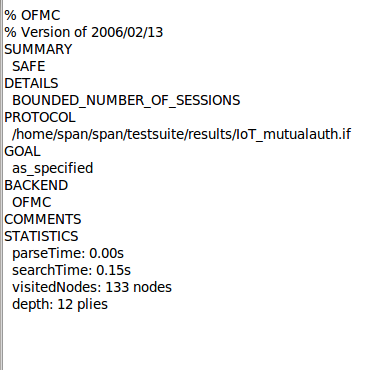}
\caption{AVISPA simulation results using OFMC (Multiple session)}
\label{ofmc1}
\end{figure}

\begin{figure}[!t]
\centering
\includegraphics[width=0.3\textwidth]{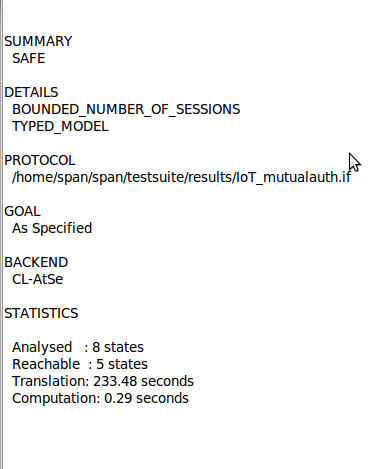}
\caption{AVISPA simulation results using ATSE}
\label{atse}
\end{figure}

\subsection{Security Analysis using Avispa Tool}
When it comes to formal verification of security protocols, AVISPA is a go-to tool \cite{armando2005avispa}. To engage in a rigorously formal security assessment of our  protocol, we harness the SPAN+AVISPA program. SPAN (security protocol animator) for AVISPA  is an indispensable tool expressly designed for CAS+ and HLPSL (high-level specification language) specifications \cite{saillard2011cas+}. Additionally, utilizing SPAN's active intruder implementation, assaults on protocols can be found and modeled interactively. AVISPA helps determine whether the security protocol is safe against active/passive attacks like MITM and message replay. 

In the process of assessing protocol security, AVISPA leverages four primary backend components. These encompass a constraint logic-based attack searcher known as CL-AtSe, a SAT-based model checker called SATMC, an on-the-fly model checker (OFMC) specialized in identifying established attack patterns, and a tree automata technique that relies on automatic approximations for the comprehensive analysis of security protocols, known as TA4SP \cite{nimmy2023novel}. The AVISPA software is not compatible with the SATMC or TA4SP back-ends, so XOR operations cannot be used.
As a result, we conduct CL-AtSE and OFMC back-ends to assess our proposed protocol safety. 
For the Authentication and Key establishment phase of our proposed protocol, we employ HLPSL, which is subsequently translated into IF for compatibility with AVISPA. Subsequently, the IF specifications are input into model checkers like OFMC and CL-ATSe for analysis. These protocols undergo examination by the backend components, all while considering the Dolev-Yao threat model. In addition, an intruder attack is simulated using parameters that are already in the public domain. CL-AtSe and OFMC both conduct searches of malicious intruders to ensure that only authorized parties can carry out the specified protocol. The information about regular sessions between the legitimate entities is then provided to the intrusive party by these back-ends. 

To show that our proposed protocol is safe, we first use a rule-based HLPSL to describe it. HLPSL is used to set up the different roles for the Client device, the Server, and the IoT device, as well as the required roles for the security goals, environment, and sessions. We check the proposed protocol's security using the model checkers OFMC and CL-AtSe, as can be seen in Figures \ref{ofmc2}-\ref{atse}. Upon executing the HLPSL scripts within AVISPA, our proposed protocol underwent assessment, ultimately receiving a favorable evaluation as "SAFE." This signifies that the suggested protocol exhibits robust protection against the Dolev-Yao threat model, specifically demonstrating resilience against message replay and Man-in-the-Middle (MITM) attacks.
\subsection{Informal Security Analysis}
In this section, we informally analyze the security of our proposed protocol.
\subsubsection{MITM Protection and Mutual Authentication}
Let's imagine that during the Authentication step, an attacker tries to intercept and modify the communication between the Client and IoT devices. An attacker intercepts the Client device's initial connection request message and impersonates it in order to send the IoT device a false connection establishment request message. The $Id_{c}$ in this bogus message may be different. The IoT device receives the attacker's fictitious connection request and interprets it as coming from the Client device. It then sends the Client device a connection establishment request message that includes both the client device's identification ($Id_{c}$) and its own identity ($Id_{p}$). The IoT device's request for connection establishment is forwarded by the Client device to the Server.
Based on the actual IoT device identification (Idp), the Server extracts the correct PUF-CRPs ($C_{p}$, $R_{p}$) from its secure database.
The Server then uses the appropriate PUF-CRPs to compute $M_{1}$, $M_{2}$, $M_{3}$, and $M_{4}$ based on the actual values of $T_{1}$ and $T_{2}$ and sends it to the Client device who forwards it to the IoT device.
At this stage, the attack is unsuccessful because the attacker lacks access to the appropriate PUF-CRPs and is unable to determine the appropriate values for $M_{1}$, $M_{2}$, $M_{3}$, and $M_{4}$. Hence, MITM attack is not possible in our proposed protocol.

Therefore, in the proposed protocol, only the legitimate IoT device can generate the response $R_{(p)}$ corresponding to challenge $C_{(p)}$ applied to its PUF, and only the legitimate Client with correct $Id_{c'}$ be able to generate the Session Key and hence our proposed solution provides mutual authentication as well.

\subsubsection{Message Replay Protection}
In our proposed protocol, we are providing protection against message replay because the attacker can't replay the messages $M_{1}$, $M_{2}$, and $M_{3}$ as they contain random numbers $T_{1}$ and $T_{2}$ that changes for each session. Also, we are not storing a set of CRPs statically in the Server, and only one CRP is initially securely stored for a single session during the enrollment phase, after which this old CRP will be updated with a new CRP for every new session via the XOR operation as seen in messages $M_{5}$ and $M_{6}$. Because the attacker doesn't have access to the random number $T_{2}$, therefore, the attacker can't replay messages $M_{5}$ and $M_{6}$.

\subsubsection{Secure Session Key agreement}
The IoT device and the Client both securely generate the same session Key, H($N_{c}$,$N_{p}$,$Id_{c'}$,$ID_{p}$), at the conclusion of the mutual Authentication phase in the proposed technique. The Session key is derived from the two random nonces, $N_{c}$ and $N_{p}$, that is freshly generated by both the Client and IoT device for every new session. 
The suggested approach can, therefore, offer a secure session Key agreement. 

\subsubsection{Key escrow}
Within our proposed protocol, it's important to note that the Server does not possess access to the Session Key established between the IoT device and the Client device. This is due to the fact that only the Client device and the IoT device are directly engaged in the generation of the Session Key. Consequently, this design feature effectively safeguards our solution against the risk of Key escrow attacks. 

\subsection{Security Comparison}

\begin{table*}[!t]
\caption{Comparison of Security Features\label{tab:table6}}
\centering
\begin{tabular}{|c||c||c||c||c||c||c|}
\hline
Features & \cite{li2020provably} & \cite{roy2022plake} & \cite{chatterjee2018building} & \cite{braeken2018puf} & \cite{alladi2020secauthuav} & Proposed\\
\hline
Mutual Authentication & \checkmark & \checkmark & \checkmark & \checkmark & \checkmark & \checkmark \\
\hline
Secure Session Key & \checkmark & \checkmark & \text{\ding{55}} & \text{\ding{55}} & \checkmark & \checkmark \\
\hline
MITM protection & \checkmark & \checkmark & \checkmark & \checkmark & \checkmark & \checkmark \\
\hline
Message Replay protection & \checkmark & \checkmark & \checkmark & \checkmark & \checkmark & \checkmark \\
\hline
No Key Escrow issue & \checkmark & \text{\ding{55}} & \text{\ding{55}} & \text{\ding{55}} & \text{\ding{55}} & \checkmark \\
\hline
\makecell{Direct Server and IoT device\\participation via gateway/internet} & \text{\ding{55}} & \checkmark & \checkmark & \checkmark & \checkmark & \text{\ding{55}} \\
\hline
Lightweight Cryptography & \text{\ding{55}} & \checkmark & \text{\ding{55}} & \text{\ding{55}} & \checkmark & \checkmark \\
\hline
\end{tabular}
\end{table*}

Table \ref{tab:table6} presents a comparative analysis of the security attributes, highlighting distinctions between our proposed protocol and other pertinent protocols.

Based on Table \ref{tab:table6}, we can see that  our proposed scheme and all the existing schemes \cite{li2020provably,roy2022plake,chatterjee2018building,braeken2018puf,alladi2020secauthuav} provide MITM Protection, Mutual Authentication, and Message Replay Protection.

The schemes \cite{chatterjee2018building,braeken2018puf} don't provide the Secure Session key that is provided in all other schemes, including our proposed scheme. 

Only \cite{li2020provably} and our proposed scheme provides protection against Key Escrow attack and don't require the active participation of the Server and the IoT device and are hence suitable for devices that use BLE, ZigBee, etc.

Finally, only \cite{roy2022plake},\cite{alladi2020secauthuav}, and our proposed scheme makes use of Lightweight Cryptography that involves only using XOR and hash operations.

Hence, it can be seen that our proposed scheme outperforms other related schemes.

\section{Experimental validation and Performance analysis}

\begin{figure}[!t]
\centering
\includegraphics[width=0.4\textwidth]{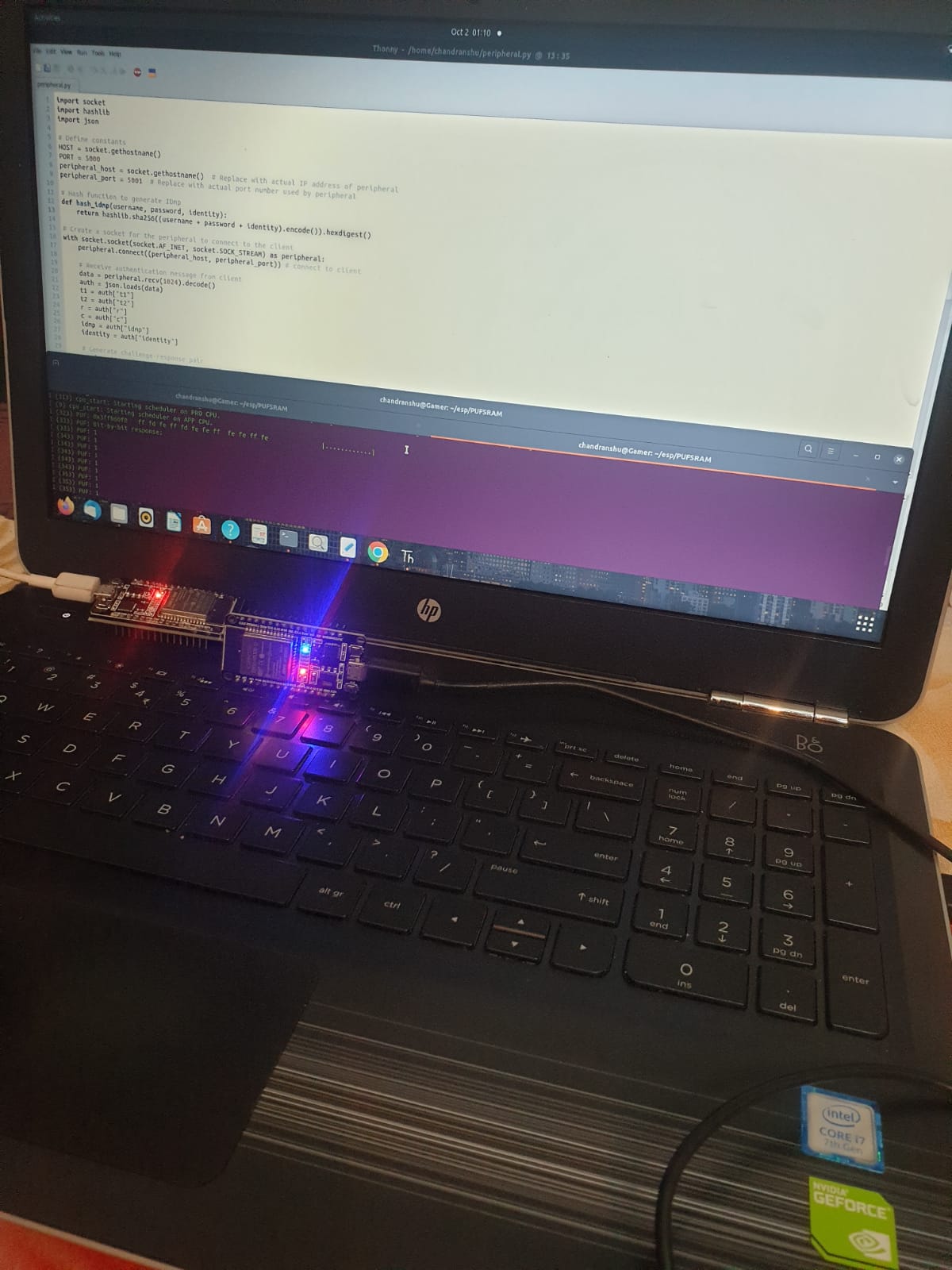}
\caption{Experimental testbed for SRAM-PUF based Authentication protocol}
\label{proto}
\end{figure}

The prototype implementation of our suggested approach is seen in Fig \ref{proto}. As an IoT device, an ESP32 microprocessor is utilized.  The local Server and client devices are both implemented on a personal computer. The PUFs may be incorporated into IoT devices as long as they have strong uniqueness and randomness to reduce the likelihood of response collision or correlation, which an attacker might exploit. The PUF may be easily chosen to satisfy any particular application needs or device resource requirements. In our instance, the internal SRAM of the ESP32 device is used to implement an SRAM-based PUF. The concept of SRAM-PUF was given by  Guajardo et al. and Holcomb et al. in \cite{guajardo2007fpga,holcomb2007initial}. In these studies, the authors discovered that the SRAM cells randomly take values of 0 or 1 during startup, and these starting values of the SRAM Cells are largely steady and unchanging after recording the starting values of SRAM cells several times.
As the starting values of SRAM cells vary amongst SRAM chips, they may serve as possible "fingerprints" for the local devices in which they are put.

Hence, we begin our implementation by reading the SRAM cells' startup values and generating a fingerprint of 96b from those SRAM cells that generate stable startup values. The addresses of the stable SRAM cells serve as PUF challenges in our suggested approach, and the corresponding fingerprint will serve as the PUF response. These addresses are chosen to take advantage of the atom-level fluctuations throughout the manufacturing process of the CMOS transistors present in SRAM that initialize the startup values of each SRAM cell to either zero or one \cite{farha2020sram}. 

Given that we are utilizing 96b as the fingerprint, we can produce a total of 96! challenges. The IoT device does not immediately communicate the starting values for these addresses after receiving the challenge. Instead, it manipulates the PUF value and uses a hashing algorithm to obscure the original SRAM cell values.  It blends the values in such a way that the final result is unique and unpredictable.

The other cryptographic modules, including hash and XOR, are loaded straight from Python libraries. The suggested protocol is built on top of the network socket for demonstration purposes.

To evaluate the effectiveness and low overhead requirements of the proposed protocol, its performance is analyzed in detail below. The performance of our proposed protocol is then compared with the existing related schemes.

\subsection{Computation Cost}

\begin{table}[!t]
\caption{EXECUTION TIME (IN MILLISECONDS) FOR ESP32\label{tab:table1}}
\centering
\begin{tabular}{|c||c|}
\hline
Operation & Average time (ms)\\
\hline
$T_{ECC}$ & 131\\
\hline
$T_{MAC}$ & 10\\
\hline
$T_{Hash}$ & 0.439\\
\hline
$T_{SED}$ & 0.259\\
\hline
$T_{XOR}$ & 0.144\\
\hline
$T_{PUF}$ & 1.9\\
\hline
\end{tabular}
\end{table}

\begin{table*}[!t]
\caption{COMPARISON OF COMPUTATION COST\label{tab:table2}}
\footnotesize
\centering
\begin{tabular}{|c||c||c||c||c|}
\hline
Schemes & Device 1 & Device 2 & Server & Total Cost\\
\hline
\cite{li2020provably} & 6$T_{h}$+7$T_{em}$+2$T_{puf}$+2$T_{xor}$ & 6$T_{h}$+7$T_{em}$+2$T_{puf}$+2$T_{xor}$  & 0 & 12$T_{h}$+14$T_{em}$+4$T_{puf}$+4$T_{xor}$\\
\hline
\cite{roy2022plake} & 3$T_{h}$+1$T_{puf}$+7$T_{xor}$ & 3$T_{h}$+1$T_{puf}$+7$T_{xor}$ & 4$T_{h}$+7$T_{xor}$ & 10$T_{h}$+2$T_{puf}$+21$T_{xor}$\\
\hline
\cite{chatterjee2018building} & 2$T_{h}$+2$T_{em}$+1$T_{puf}$+7$T_{hp}$+1$T_{bp}$ & 2$T_{h}$+2$T_{em}$+1$T_{puf}$+7$T_{hp}$+1$T_{bp}$ & 10$T_{h}$+4$T_{em}$+6$T_{hp}$+2$T_{bp}$ & 14$T_{h}$+8$T_{em}$+2$T_{puf}$+20$T_{hp}$+4$T_{bp}$\\
\hline
\cite{braeken2018puf} & 6$T_{h}$+3$T_{em}$+2$T_{puf}$+1$T_{xor}$ & 6$T_{h}$+3$T_{em}$+2$T_{puf}$+1$T_{xor}$  & 10$T_{h}$+1$T_{em}$+2$T_{xor}$  & 22$T_{h}$+7$T_{em}$+4$T_{puf}$+4$T_{xor}$\\
\hline
\cite{alladi2020secauthuav} & 4$T_{h}$+1$T_{puf}$+8$T_{xor}$ & 5$T_{h}$+1$T_{puf}$+8$T_{xor}$ & 9$T_{h}$+8$T_{xor}$ & 18$T_{h}$+2$T_{puf}$+24$T_{xor}$\\
\hline
Proposed protocol & 5$T_{h}$+1$T_{puf}$+7$T_{xor}$ & 2$T_{h}$+2$T_{xor}$ & 4$T_{h}$+5$T_{xor}$ & 11$T_{h}$+1$T_{puf}$+14$T_{xor}$\\
\hline
\end{tabular}
\end{table*}

\begin{table}[ht!]
\caption{COMPARISON OF COMMUNICATION COST\label{tab:table3}}
\begin{tabular}{|p{1.2cm}||p{1.2cm}||p{1.2cm}||p{1.2cm}||p{1.2cm}|}
\hline
{Schemes} & {Device 1 (bits)} & {Device 2 (bits)} & {Server (bits)} & {Total Cost (bits)}\\
\hline
\cite{li2020provably} & 2080  & 2080  & 0  & 4160 \\
\hline
\cite{roy2022plake} &  768  & 768  & 3072  & 4608 \\
\hline
\cite{chatterjee2018building} & 2112  & 2048  & 5696  & 9856 \\
\hline
\cite{braeken2018puf} & 864  & 768  & 4672  & 6304 \\
\hline
\cite{alladi2020secauthuav} & 1568  & 1824  & 1536  & 4928 \\
\hline
Proposed protocol & 1536  & 2816  & 1536  & 5888 \\
\hline
\end{tabular}
\end{table}

To calculate the computation time required to perform different cryptographic operations used in different Authentication and Key Exchange protocols, we utilized the ESP-32 board, which has a 240 MHz Tensilica Xtensa LX6 CPU, 520 KB of SRAM, and 4 MB of Flash memory running on Thonny IDE. The code has been written in MicroPython using the inbuilt libraries. Table \ref{tab:table1} provides the experiment's findings for the ESP32 configuration. 
The various operations that are compared include $T_{XoR}$, $T_{ECC}$, $T_{Hash}$, $T_{PUF}$, $T_{MAC}$, and $T_{SED}$ to calculate the execution times needed for "XOR operation," "elliptic curve scalar point multiplication," "hash function" (e.g., SHA-256), "MAC based encryption," and "symmetric Key encryption/decryption." 

As can be seen from Table \ref{tab:table1}, the ECC-based public-key cryptosystem is outperformed by the MAC-based symmetric cryptosystem in terms of the execution time taken. In our proposed protocol, we have completely removed the ECC-based protocols and the MAC computation's computational barrier. By only using the hash and XOR operations along with the PUF, we have reduced the computation cost in terms of time taken to a large extent.

In Table \ref{tab:table2}, we have shown the comparison of the computational cost of our proposed protocol with related schemes. Let $T_{xor}$, $T_{em}$, $T_{h}$, $T_{puf}$, $T_{mac}$, $T_{hp}$, and $T_{bp}$, stand for the total number of xor operations, point multiplication over ECC operations, general hash operations, PUF operations, MAC operations, map-to-point hash operations, and bilinear pairing operations, respectively during the Authentication phase of the protocols. 

Based on Table \ref{tab:table2}, it is clear that our proposed protocol and \cite{roy2022plake,alladi2020secauthuav} are the most efficient and cost-effective options based on the lightweight cryptographic operations involved and outperforms other related protocols. The map-to-point hash, point multiplication, and bilinear pairing operations used in \cite{chatterjee2018building} have high computational complexity and thus increase the cost of running the protocols. \cite{li2020provably,braeken2018puf} too rely on point multiplication, which also has high computational complexity when compared with the xor and hash operations used in other protocols. Our proposed protocol even outperforms the related lightweight schemes \cite{roy2022plake,alladi2020secauthuav} since it utilizes less XOR and PUF operations. 

\subsection{Communication Cost}
In this section, we will compare the communication overhead of our proposed protocol with other related schemes during the Authentication and Key Exchange phases of the protocols.

We have assumed that 256 bits be used for random numbers, the output of hash and MAC functions, private Keys, challenges, and responses from the PUF, while 32 bits be used for device identifiers and timestamps. Points on an elliptic curve are 512 bits in length and of the form (Px, Py), where Px and Py stand for the x and y coordinates, and both are 256 bits long, respectively. Moreover, we have only considered the Authentication and Key Exchange phase of the protocol and not the initial request-response messages exchanged between the devices/Servers where they exchange device IDs and global parameters related to ECC. Including these request-response messages will significantly increase the communication cost of \cite{li2020provably,chatterjee2018building,braeken2018puf} without impacting \cite{roy2022plake,alladi2020secauthuav} and our proposed scheme.

Table \ref{tab:table3} displays the results of a comparison of the communication cost. The proposed protocol's communication cost is clearly superior to those of the related schemes \cite{chatterjee2018building,braeken2018puf}. The scheme in \cite{li2020provably} shows less communication cost than our proposed scheme, but it emits the communication cost of the initial request-response messages exchanged between the devices/Servers.

\section{Conclusion}
To solve the problem of Authentication and Key exchange in IoT devices, many schemes based on PKI, IBE, PUF, and Symmetric key Cryptography have been proposed. Based on our understanding, we found out that PUF-based and Symmetric key Cryptography based Authentication and Key exchange schemes are the right choice for these resource-constrained IoT devices since the performance barrier caused by resource-intensive PKI and IBE schemes is substantially removed by PUF-based and Symmetric key Cryptography based schemes.

With the use of PUFs, this research has developed a lightweight, robust, and secure Authentication and Key Exchange protocol for resource-constrained IoT devices. 
The protocol's computing performance is greatly improved by using PUF technology, as the IoT devices no longer need to store secrets in their memory and can Authenticate themselves without direct communication via the internet or gateway with the Server during the Authentication phase and hence our proposed scheme can be utilized by the devices that use BLE, Zigbee, etc. 
The Server doesn't need to store the set of CRPs in its database, and only a single CRP per device is required to be stored, which helps secure our protocol against the mathematical modeling of PUF. Using the formal verification tool Avispa, it has been confirmed that our protocol provides a secure secret session key, provides protection against replay attacks, and provides the desired mutual Authentication. 

The proposed protocol is limited to a maximum communication overhead of 4352 bits, and the computation cost of 11$T_{h}$+1$T_{puf}$+14$T_{xor}$ which when compared with other related schemes confirms that it is lightweight.


\bibliographystyle{model1-num-names}

\bibliography{cas-refs}


\end{document}